\documentclass[prl,twocolumn,english,showpacs]{revtex4-1}
\usepackage{amssymb}
\usepackage{color}
\usepackage{times}
\usepackage{graphicx}
\usepackage{bbold}
\usepackage{bm}
\usepackage{amsmath}
\usepackage{amsfonts}
\usepackage{amssymb}
\usepackage{braket}
\begin{document}

\title{Thermodynamics vs. local density fluctuations in the metal/Mott-insulator crossover}

\author{J. H. Drewes$^{1}$,  E. Cocchi$^{1,2}$,  L. A. Miller$^{1,2}$, C. F. Chan$^1$, D. Pertot$^{1}$, F. Brennecke$^{1}$, and M. K{\"o}hl$^{1}$}

\affiliation{$^1$Physikalisches Institut, University of Bonn, Wegelerstrasse 8, 53115 Bonn, Germany\\
$^2$Cavendish Laboratory, University of Cambridge, JJ Thomson Avenue, Cambridge CB3 0HE, United Kingdom}

\begin{abstract}
The crossover between a metal and a Mott insulator  leads to a localization of fermions from delocalized Bloch  states to localized states.  We experimentally study this crossover using fermionic atoms in an optical lattice by measuring thermodynamic and local (on--site) density correlations. In the metallic phase  at incommensurable filling we observe the violation of the local fluctuation--dissipation theorem indicating that the thermodynamics cannot be explained by local observables. In contrast, in the Mott-insulator we observe the convergence of local and thermodynamic fluctuations  indicating the absence of long--range density-density correlations. 
\end{abstract}
\maketitle


Understanding strongly-correlated quantum many-body systems remains a challenge both for experiment and theory. In particular, the interplay of interactions, kinetic energy and dimensionality is complex and governs the occurrence and properties of low-temperature quantum phases. A paradigmatic example of a strongly-correlated many-body problem is that of interacting spin-1/2 fermions on a periodic lattice. Depending on the Hamiltonian parameters,  different quantum phases are realized. For example, for weak interactions and low filling  of the  lattice, the fermions delocalize into Bloch waves and constitute a metallic state with finite charge compressibility. In contrast, for strong repulsive interactions at half filling the fermions form a Mott insulator  which  occurs  when the kinetic energy can no longer  overcome the energy gap due to repulsive on-site interactions  \cite{Imada1998}. It is believed that  Mott physics is responsible for many features of strongly correlated materials, for example in the  cuprate superconductors \cite{Dagotto1994,Lee2006}. In recent years, ultracold atoms in optical lattices have emerged as a versatile and powerful platform to study the physics of spin-1/2 fermions in a lattice and thus emulate strongly-correlated materials.

A key requirement to understanding strongly-correlated phases is the establishment of a link between the microscopic physics and macroscopic observables. In a Mott insulator,  on the microscopic level, strong repulsive interactions lead to a suppression of doubly-occupied lattice sites and to a localization of atoms at the lattice sites. This goes hand in hand with the disappearance of long-range correlations of the density fluctuations. On a macroscopic level the Mott insulator exhibits a vanishing  compressibility. This thermodynamic quantity can be inferred from measurements averaging over  ensembles \cite{Schneider2008,Duarte2015,Cocchi2016}. Conceptually, the link between thermodynamic and microscopic observables is provided by  the fluctuation--dissipation theorem \cite{Kubo1966}.  The fluctuation--dissipation theorem quantifies how the presence of intrinsic microscopic fluctuations is linked to the response of a system to a (weak) external perturbation such as a variation of the chemical potential $\mu$. Fluctuation--dissipation theorems can be derived for a variety of linked thermodynamic/microscopic observables.
In this work, we study the crossover  between a metal and a Mott-insulator which is observable in density (charge) ordering. Hence we concentrate on the relation between density fluctuations and the isothermal  compressibility $\kappa$.
The corresponding fluctuation--dissipation theorem states that the isothermal compressibility $\kappa(\bm{r})$  and the density correlations at location $\bm{r}$ are linked by \cite{Zhou2011}
\begin{equation}
\kappa(\bm{r}) = \frac{1}{k_BT}\int d\bm{r}^\prime \left[\braket{\rho(\bm{r})\rho(\bm{r}^\prime)}-\braket{\rho(\bm{r})}\braket{\rho(\bm{r}^\prime)}\right].
\end{equation}
Here, $\rho(\bm{r})$ is the density, $T$ is temperature and $k_B$ is Boltzmann's constant. This formulation of the fluctuation--dissipation theorem provides us with two key insights. First, the proportionality factor between the thermodynamic variable $\kappa$ and the microscopic density correlations is the temperature. This property has been employed, for example, to conduct fluctuation--based thermometry \cite{Muller2010,Sanner2010} when compressibility and density-fluctuations can be independently measured. Second, in principle, density fluctuations at all length scales contribute to the thermodynamic compressibility. This effect has yet to be fully explored. Previous work has focussed on column-integrated densities \cite{Muller2010,Sanner2010} or on local measurements \cite{Omran2015}, such that long-range density correlations were either integrated out or not detected.

\begin{figure*}
 \includegraphics[width=\textwidth,clip=true]{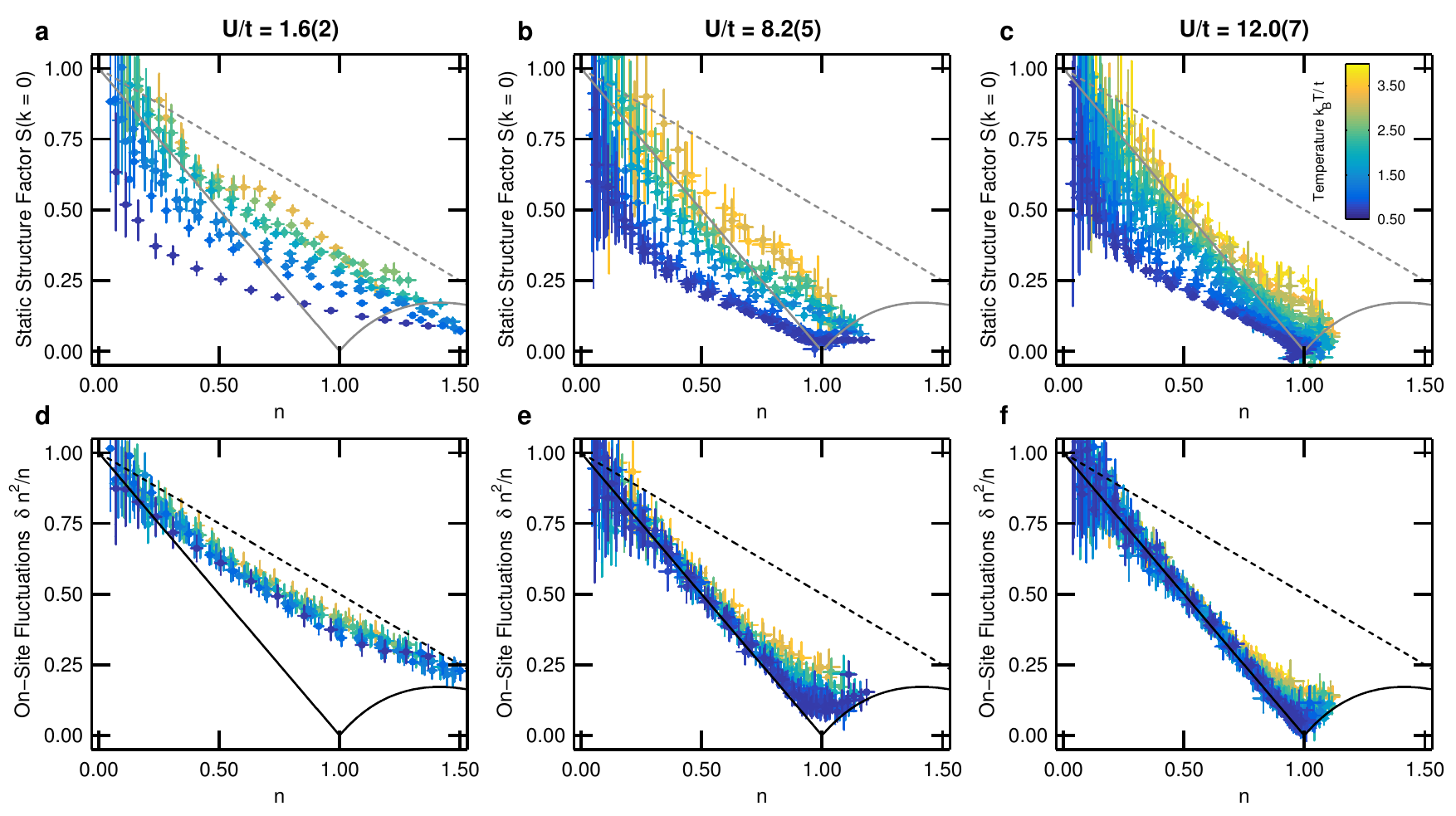}
 \caption{Comparison of the thermodynamic and local density fluctuations in the two-dimensional Hubbard model. Left column (a,d): $U/t=1.6(2)$, middle column (b,e): $U/t=8.2(5)$, right column (c,f): $U/t=12.0(7)$. Top row: static structure factor $S(k=0)$  measuring density fluctuations in thermodynamically large volumes; bottom row: density fluctuations at a single site of the optical lattice. Dashed line: on-site fluctuations in the non-interacting limit. Solid line: on-site fluctuations in the infinite-interactions limit. Temperature is encoded as color.}
\end{figure*}

The theoretical description of the interacting gas on a lattice  is given by the Hubbard model. This model  considers the two elementary processes of tunneling between neighboring lattice sites with amplitude $t$ and on-site interaction between two fermions of opposite spin  with strength $U$. In a single-band approximation the Hubbard Hamiltonian reads
\begin{eqnarray}
\hat{H}=-t\sum_{\braket{i,j},\sigma}\left(\hat{c}_{i\sigma}^\dag \hat{c}_{j\sigma} +h.c.\right) +U\sum_{i}\hat{n}_{i\downarrow}\hat{n}_{i\uparrow}.
\end{eqnarray}
Here $\hat{c}_{i\sigma}$ ($\hat{c}_{i\sigma}^\dag$) denotes the annihilation (creation) operator of a fermion on lattice site $i$ in spin state $\sigma=\{\uparrow, \downarrow\}$, the bracket $\braket{,}$ denotes the restricted sum over nearest neighbours, and $\hat{n}_{i\sigma}=\hat{c}^\dag_{i\sigma}\hat{c}_{i\sigma}$ is the number operator. The Mott insulator forms at half filling, i.e. for $n= \braket{\hat{n}_{i\uparrow}}+\braket{\hat{n}_{i\downarrow}}=1$, and $U\gg t, k_BT$. It is characterized by an occupation of one particle per lattice site and a gap against density excitations (i.e. doubly-occupied lattice sites)  of order $U$.

The Hubbard model can be realized using ultracold atoms in an optical lattice. Earlier work investigating  the Mott insulator  studied the global disappearance of doubly occupied sites \cite{Joerdens2008,Joerdens2010,Taie2012} and the response to an external compression \cite{Schneider2008}. However, ultracold atoms are usually confined by an external trapping  potential $V(\bm{r})$ leading to a spatially varying density distribution $\rho(\bm{r})$. Therefore, different quantum phases  coexist in different regions of the trap. In order to take advantage of the coexistence,  spatially-resolved imaging techniques have recently been implemented for Fermi gases in optical lattices \cite{Duarte2015,Haller2015,Greif2015,Hofrichter2016,Edge2015,Cocchi2016,Omran2015,Cheuk2016}.

In our experiment, we prepare a spin-balanced  quantum degenerate mixture of the two lowest hyperfine states $\ket{F=9/2,m_F=-9/2}$ and $\ket{F=9/2,m_F=-7/2}$ of fermionic atoms of the isotope $^{40}$K \cite{Frohlich2011,Cocchi2016}. We load the quantum gas into an anisotropic, three-dimensional optical lattice in which  tunneling is suppressed along the vertical direction by means of a  high lattice depth. Hence, the dynamics is restricted to two-dimensional planes within which we choose a lattice depth of $5.2\leq V_{xy}\leq 6.6$\,$E_{\text{rec}}$, where $E_{\text{rec}}=\hbar^2 \pi^2/(2 m a^2)$ denotes the recoil energy, $a=532\,$nm is the lattice period, and $m$ is the atomic mass. The Hubbard interaction parameter $U$ is controlled by utilizing a Feshbach resonance near 202\,G which provides us with access to the parameter range from weak to strong interactions $1.6\leq U/t\leq 12.0$. Additionally, the temperature of the gas is adjusted by heating either due to periodic modulation of the trapping potential followed by a thermalization time, or by holding the atoms in the lattice and subjecting them to heating from inelastic interaction with lattice laser light.  We prepare equilibrium systems with well-defined parameters $t$, $U$, and $k_BT$ and detect the density distribution in a single two-dimensional layer of the optical lattice \cite{Cocchi2016}. By combining radio-frequency spectroscopy and absorption imaging we separately detect the in-situ density distributions of singly-occupied lattice sites,  $\braket{\hat{n}_{i\uparrow}-\hat{n}_{i\uparrow} \hat{n}_{i\downarrow}}$, and  doubly-occupied lattice sites (``doubles''), $\braket{\hat{n}_{i\uparrow} \hat{n}_{i\downarrow}}$ in a single measurement. Our technique gives direct access to the equation of state \cite{Ho2010} $n(\mu)$, where $\mu$ denotes the chemical potential. We perform thermometry by fitting the measured data with either numerical linked cluster expansion calculations of the two-dimensional Hubbard model \cite{Khatami2011} or the ideal-Fermi gas density distribution in the low-density wings of the cloud. Finally, we derive the compressibility $\kappa$ by  numerical differentiation of the interpolated experimental data as $\kappa(\mu)=\frac{\partial \rho}{\partial\mu}$, for details see \cite{Cocchi2016}.

We first investigate the relationship between thermodynamic and local fluctuations. Thermodynamically, the number fluctuations of a gas in thermal equilibrium follow from the fluctuation-dissipation theorem (1) as
\begin{equation}
\frac{\delta N^2}{N}=\frac{k_B T \kappa}{\rho}=S(k=0).
\end{equation}
Here,  $S(k)$ is the static structure factor at wave vector $k$ and relation (3) holds for infinitely large volumes \cite{Klawunn2011b}. Qualitatively, one discriminates three different cases $S(k=0)>1$ is termed bunching, $S(k=0)=1$ is  classical Poissonian shot noise, and $S(k=0)<1$ is anti--bunching, which is expected for a degenerate Fermi gas. In Figures 1a-c, we show the static structure factor $S(k=0)$ derived from the measurement of compressibility and density for different temperatures and interactions. Generally, we find that for high-filling and low temperature the structure factor is well below 1, which signals non-classical behaviour. For weak interactions, $U/t=1.6$, this displays anti-bunching due to the Pauli exclusion principle according to which at most one fermion per spin state can occupy each lattice site. For strong interactions, $U/t=12.0$, the structure factor is even more suppressed by the strong repulsive interaction between atoms. For high temperatures, the structure factor increases due to the thermal contribution.

We now compare the structure factor with the local fluctuations $\delta n^2/n$ on one lattice site. The on--site density fluctuations $\delta n^2=\braket{\hat{n}_i^2}-\braket{\hat{n}_i}^2$ follow from the measured densities using the fermionic anti--commutation relations and using  $\braket{\hat{n}_{i\uparrow}}=\braket{\hat{n}_{i\downarrow}}$ as
\begin{equation}
\delta n^2=2\braket{\hat{n}_{i\uparrow}}-4\braket{\hat{n}_{i\uparrow}}^2+2 \braket{\hat{n}_{i\uparrow} \hat{n}_{i\downarrow}}.
\end{equation}
This expression provides us with a powerful analysis tool:  one can determine the on-site density fluctuations from a measurement of the density distributions of the single-- {\it and} double--occupancy of the lattice sites. When applying relation (4), we assume that the global density does not vary significantly over the extent of the spatial resolution of our imaging system. We have determined our imaging resolution (FWHM of the autocorrelation peak) of $1.2(1)\,\mu$m and we can safely assume that there is no density variation from the harmonic trapping potential on such short length scales. In Figures 1d-f we plot the local density fluctuations for the same data as in Figure 1a-c. For the ideal Fermi gas on a lattice the local fluctuations follow a binomial distribution which is imposed by Pauli's exclusion principle. Specifically, in the non-interacting limit either zero or one fermion per spin state can occupy each lattice site with the mean given by the particle number per site $n$, so the fluctuations follow a binomial distribution $\delta n^2/n= 1-n/2$  (dashed line). The measured local fluctuations for $U/t=1.6$ reproduce the ideal Fermi gas prediction very well, with an additional small suppression of the fluctuations which we attribute to the finite interaction strength. In the limit of infinite interactions and $n \leq 1$, either zero or one fermion occupy one lattice site and hence $\delta n^2/n = 1-n$. For $n>1$ we find in the same limit $\delta n^2/n = 3-n-2/n$. The on-site fluctuations in the strongly-interacting case are insensitive to temperature unless $k_B T \sim U$ for which thermally induced double--occupancies contribute to the fluctuations. 

Generally, we find that the on-site density fluctuations $\delta n^2/n$ for a given filling and temperature are higher than the fluctuations $\delta N^2/N$ of the same data set in the thermodynamic limit.  This shows that the  thermodynamic fluctuations contain a  non-local contribution from density-density correlations on different length scales. This contribution prohibits the application of a local fluctuation--dissipation theorem  and often inhibits the determination of thermodynamic quantities by local quantities. In order to understand the difference between local and thermodynamic fluctuations, we transcribe the fluctuation--dissipation theorem (1) to a lattice with discrete sites labelled by the index $i$
\begin{equation}
\kappa=\frac{1}{a^2k_B T}\left[\delta n^2+\sum_{j\neq i}\left(\braket{\hat{n}_i\hat{n}_j}-\braket{\hat{n}_i}\braket{\hat{n}_j}\right)\right].
\end{equation}
Here we have separated the local fluctuation $\delta n^2$ from the non-local density correlations. On the microscopic level, the violation of the local fluctuation--dissipation theorem\cite{Duchon2012,Fang2011} is rooted in the spatial correlations of the density fluctuations and hence is governed by the nature of the underlying quantum state. In a perfectly localized state, such as a band insulator at zero temperature, the off-site correlations are zero  since the correlation function factorizes $\braket{\hat{n}_i\hat{n}_j}= \braket{\hat{n}_i}\braket{\hat{n}_j}$.  In contrast, for a delocalized state at zero temperature, such as a Fermi gas in a partially filled Bloch band, the compressibility cannot be described by local fluctuations alone since the off-site correlations are not negligible. At finite temperature, however, the thermal correlation length caps the range of density correlations and for a classical gas the correlations are limited to on-site fluctuations. In order to highlight this effect, we plot  in Figure 2 the  non-local density correlation $\delta n^2_\text{n.l.}=a^2\kappa k_BT-\delta n^2$ as a function of temperature and lattice filling. For low filling, $n\lesssim 0.4$, the results  for all interaction strengths agree very well with the non-interacting Fermi gas in a two-dimensional square lattice with nearest-neighbour tunneling (solid line). This shows that at low filling the atoms delocalize in the lattice irrespective of the explored interaction strength and that for low temperatures the delocalization gives rise to non-local density correlations which significantly affect the compressibility. At $n \geq 0.5$ we observe the onset of interaction effects in the deviations from the ideal Fermi gas theory. While the weakly-interacting data (blue) remain closer to the ideal Fermi gas prediction, the strongly-interacting data  exhibit a suppression of long--range density correlations. Finally, at half-filling,  we observe a clear distinction between the two cases. For example, the $U/t=12$ data show an almost complete suppression of off-site density correlations, while the $U/t = 1.6$ data are still close to the free Fermi gas expectation. The latter is a signature for the atoms having formed a localized Mott insulator, the thermodynamics  of which in the charge sector is entirely described by local quantities.

\begin{figure}
 \includegraphics[width=.9\columnwidth,clip=true]{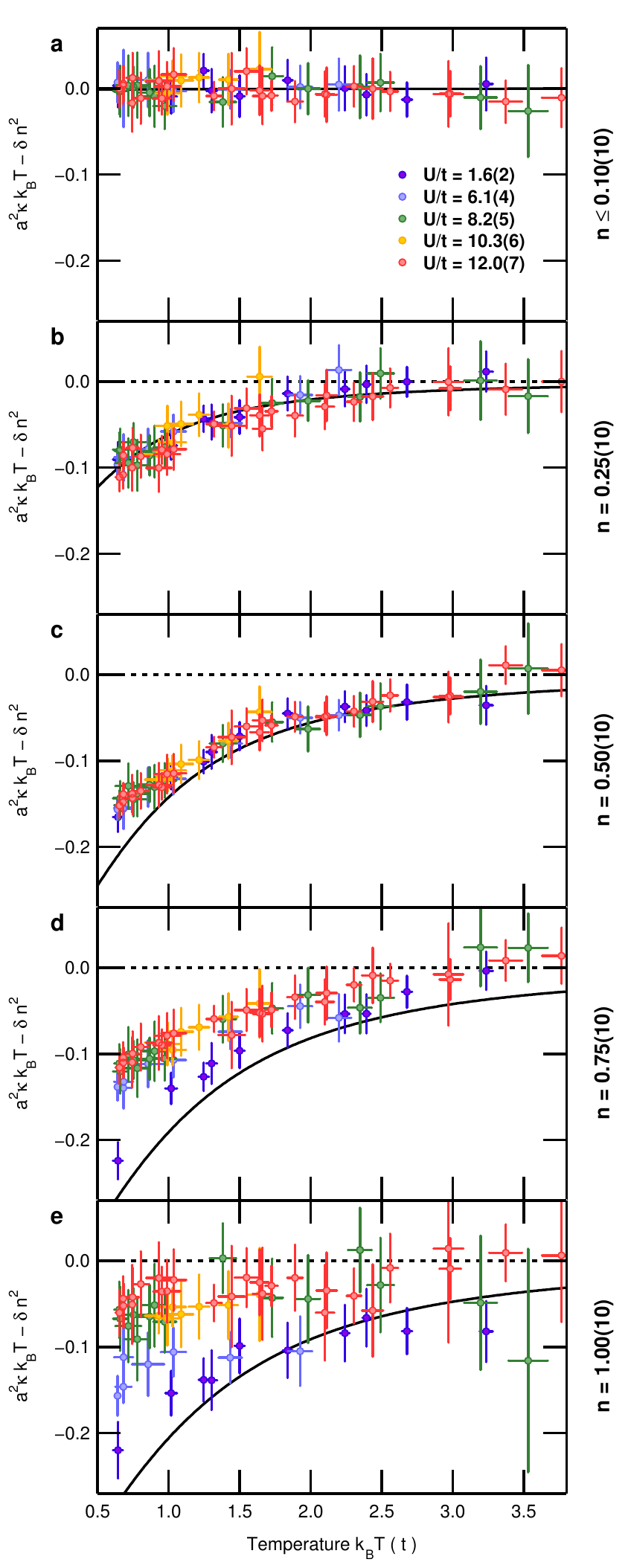}
 \caption{Nonlocal density-density correlations for different lattice occupations $n$. The solid line is the theoretical expectation of the ideal Fermi gas in a two-dimensional lattice. Color code: purple: $U/t=1.6(2)$, light blue: $U/t=6.1(4)$, green $U/t=8.2(5)$, yellow: $U/t=10.3(6)$, red: $U/t=12.0(7)$.}
\label{fig3}
\end{figure}

Finally, in Figure 3 we show, for the lowest temperatures achieved at each interaction strength, the residual non-local density fluctuations $\delta n^2_\text{n.l.}$ as a function of the atom number per site $n$. The data highlight  that  at low filling the off-site density correlations are essentially independent of interaction strength and equal to the ideal Fermi gas prediction. Upon approaching half-filling, the off-site correlations are highly suppressed for the strongly-interacting gas and signal the onset of a Mott insulator. 

\begin{figure}
 \includegraphics[width=.9\columnwidth,clip=true]{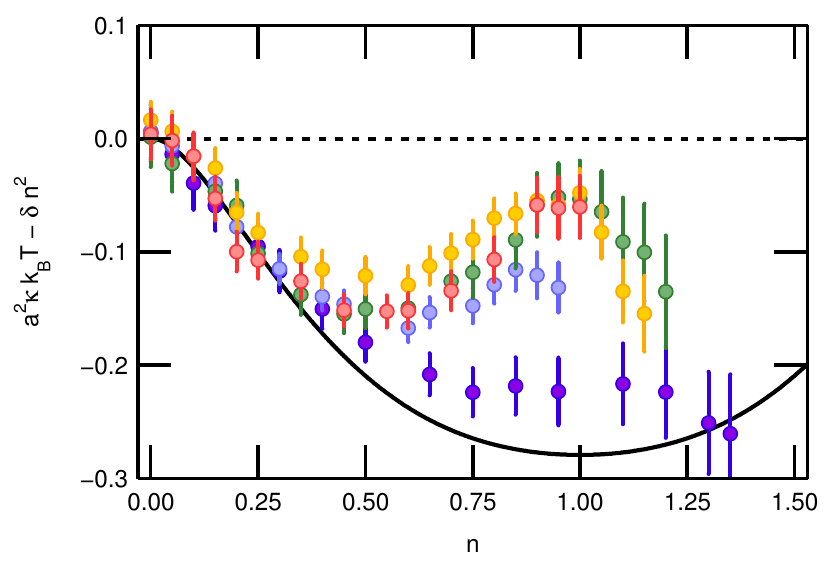}
 \caption{Non-local density-density correlations for the lowest temperatures. Color code: purple: $U/t=1.6(2)$, light blue: $U/t=6.1(4)$, green $U/t=8.2(5)$, yellow: $U/t=10.3(6)$, red: $U/t=12.0(7)$. The line shows the prediction of the ideal Fermi gas on a square lattice for $k_BT/t=0.65$.}
\label{fig3}
\end{figure}

The technique presented here could be beneficially applied to other complex many-body phenomena, such as the localization in disordered systems and quantum phase transitions in strongly correlated samples\cite{Duchon2012}. Very often, the precise determination of the transition point is hindered by the overwhelming signal from the on-site correlations. Focussing on the comparison between local and thermodynamic quantities will lead to a much improved characterization of crossovers and phase transitions in correlated many-body systems.

We thank C. Kollath for discussion and N. Wurz for experimental assistance. The work has been supported by DFG, the Alexander-von-Humboldt Stiftung, EPSRC and ERC.


\end{document}